\documentclass[11pt]{article}
\def\mainmanuscript{}
\usepackage[preprint]{acl}
\usepackage{times}
\usepackage{latexsym}
\usepackage[T1]{fontenc}
\usepackage[utf8]{inputenc}
\usepackage{microtype}
\usepackage{inconsolata}
\usepackage{graphicx}
\usepackage{xspace}
\usepackage{array}
\usepackage{amssymb}
\usepackage{multirow}
\usepackage{subcaption}
\usepackage{booktabs}
\usepackage{xcolor}
\usepackage[cachedir=minted-cache]{minted}

\newlength{\agentH}
\newlength{\agentWa}
\newlength{\agentWb}
\newlength{\agentWc}
\newcommand{\framework}{\textsc{ExploreRepair}\xspace}
\newcommand{\Description}[1]{}

\newcommand{\change}[1]{\textcolor{blue}{#1}}
\newcommand{\mynote}[2]{%
  \fbox{\bfseries\sffamily\scriptsize#1}%
  {\small$\blacktriangleright$\textsf{\emph{#2}}$\blacktriangleleft$}}%
\newcommand{\todoblue}[1]{\textcolor{blue}{\textbf{#1}}}
\newcommand{\todobrown}[1]{\textcolor{brown}{\textbf{#1}}}
\newcommand{\todored}[1]{\textcolor{red}{\textbf{#1}}}
\newcommand{\todoolive}[1]{\textcolor{olive}{\textbf{#1}}}
\newcommand{\manh}[1]{\mynote{Manh}{\todoblue{#1}}}
\newcommand{\hj}[1]{\mynote{HJ}{\todored{#1}}}
\newcommand{\iv}[1]{\mynote{Ivana}{\todobrown{#1}}}
\newcommand{\rw}[1]{\mynote{Ratna}{\todoolive{#1}}}

\begin{document}

\title{ExploreRepair: Evidence-Specialized Agents for Vulnerability Repair under Information Constraints}

\author{
  \textbf{Duc Manh Tran\textsuperscript{1}},
  \textbf{Ratnadira Widyasari\textsuperscript{2}},
  \textbf{Ivana Clairine Irsan\textsuperscript{3}},
  \textbf{Huihui Huang\textsuperscript{3}},
\\
  \textbf{Ting Zhang\textsuperscript{4}},
  \textbf{David Lo\textsuperscript{3}},
  \textbf{Rahul Gopinath\textsuperscript{1}},
  \textbf{Hong Jin Kang\textsuperscript{1}}
\\
  \textsuperscript{1}The University of Sydney, Australia,
  \textsuperscript{2}University of Victoria, Canada,
\\
  \textsuperscript{3}Singapore Management University, Singapore,
  \textsuperscript{4}Monash University, Australia
}

\maketitle
\ifdefined\mainmanuscript
\else
\documentclass[11pt]{article}
\usepackage[review]{acl}
\usepackage{times}
\usepackage{latexsym}
\usepackage[T1]{fontenc}
\usepackage[utf8]{inputenc}
\usepackage{microtype}
\usepackage{inconsolata}
\usepackage{graphicx}
\usepackage{xspace}
\usepackage{array}
\usepackage{multirow}
\usepackage{subcaption}
\usepackage{booktabs}
\usepackage{xcolor}
\graphicspath{{images/}{../images/}}
\newlength{\agentH}
\newlength{\agentWa}
\newlength{\agentWb}
\newlength{\agentWc}
\newcommand{\framework}{\textsc{ExploreRepair}\xspace}
\newcommand{\Description}[1]{}
\newcommand{\manh}[1]{}
\newcommand{\hj}[1]{}
\newcommand{\iv}[1]{}
\newcommand{\rw}[1]{}
\begin{document}
\fi

\begin{abstract}
Automated vulnerability repair must often operate with incomplete vulnerability information: developers may lack an accurate vulnerability classification or a provided proof-of-vulnerability (PoV) test, yet still need to produce a repair.
Without detailed information, a key challenge is to explore multiple plausible interpretations of the vulnerability before committing to a repair. 
To address this challenge, we propose \framework, a vulnerability repair framework that reconciles the vulnerability observations and repair implementations from a team of exploration agents.
We first collect these observations and repair implementations from three agents, each tasked with exploring a specific dimension of the vulnerability.
Then, a Lesson Aggregator and a Patch Aggregator independently synthesize repairs from these observations and concrete implementations before a final Aggregator reconciles their proposals without benchmark feedback.
On PatchEval, \framework achieves repair success rates of 39.6\% and 71.6\% with GPT-5.6-Luna and DeepSeek-V4.1-Flash, respectively, exceeding EvoRepair, a state-of-the-art vulnerability repair approach, by 16.0 and 25.4 percentage points.
On VulnLoc+, it achieves  success rates of 51.4\% and 62.9\%, respectively, exceeding LoopRepair's variant without PoV tests by 20.0 and 17.1 percentage points.
\end{abstract}

\ifdefined\mainmanuscript
\else
\bibliography{refs}
\end{document}
\fi

\ifdefined\mainmanuscript
\else
\documentclass[11pt]{article}
\usepackage[review]{acl}
\usepackage{times}
\usepackage{latexsym}
\usepackage[T1]{fontenc}
\usepackage[utf8]{inputenc}
\usepackage{microtype}
\usepackage{inconsolata}
\usepackage{graphicx}
\usepackage{xspace}
\usepackage{array}
\usepackage{multirow}
\usepackage{subcaption}
\usepackage{booktabs}
\usepackage{xcolor}
\graphicspath{{images/}{../images/}}
\newlength{\agentH}
\newlength{\agentWa}
\newlength{\agentWb}
\newlength{\agentWc}
\newcommand{\framework}{\textsc{ExploreRepair}\xspace}
\newcommand{\Description}[1]{}
\newcommand{\manh}[1]{}
\newcommand{\hj}[1]{}
\newcommand{\iv}[1]{}
\newcommand{\rw}[1]{}
\begin{document}
\fi

\section{Introduction}
\label{sec:intro}

Automated vulnerability repair (AVR) focuses on removing security flaws while preserving intended program behavior~\cite{gao2021beyond,zhang2022program,zhou2024out,nong2025appatch,ye2026well, li2026vulkey, han2026symradar}.
Recent approaches depend on complete vulnerability information, 
such as detailed vulnerability descriptions or proof-of-vulnerability (PoV) tests, to guide repair.
These artifacts help characterize the flaw and guide patch generation, 
for example by providing feedback for validating and refining candidate repairs.

In practice, developers often have to fix vulnerabilities without access to critical vulnerability information, as it is frequently missing or unavailable~\cite{guo2022detecting,sun2023aspect}. 
For example, only 4.1\% of 75,807 CVEs had an available public PoV within one year of disclosure~\cite{householder2020historical}.
Nevertheless, state-of-the-art AVR approaches, such as EvoRepair and ContraFix,
depend on information only available from a complete vulnerability analysis, including PoVs and oracle feedback, to guide repair~\cite{hu2026evorepair,liu2026contrafix}.

In this study, we investigate vulnerability repair without complete information.
This setting reflects real-world practice, in which such information is frequently missing, especially for newly discovered vulnerabilities whose fixes are urgently required.
To compensate for the loss of complete vulnerability information, a repair method has to explore the codebase to \textbf{gather evidence 
while avoiding premature commitment to an incorrect repair hypothesis.}

The need to \textbf{simultaneously consider  heterogeneous evidence types} makes this task difficult. 
First, the vulnerable code location provides incomplete \emph{structural evidence}, e.g., 
information about how inputs can reach the code, and parallel repair sites may lie elsewhere in the repository~\cite{tang-etal-2025-synfix}.
Second, source code provides incomplete \emph{semantic evidence}, e.g., the relevant API contract and intended behavior must be reconstructed. 
Third, the absence of a ground-truth PoV removes \emph{executable evidence} that demonstrates the vulnerable behavior.

To this end, we propose \framework, a vulnerability repair framework with two phases, exploration and aggregation.
During \emph{exploration}, three independent, \textbf{evidence-specialized agents} investigate repository structure, infer developer intent, or construct and execute a PoV. 
Each agent returns a candidate patch and the trajectory of its exploration.

During \emph{aggregation}, the agents' trajectories and patches are \textbf{hierarchically aggregated} to leverage their complementary information and avoid committing to a single dimension of analysis.
First, we use an LLM to distill the agents' trajectories and candidate patches into lessons that capture their vulnerability observations and supporting evidence.
Next, the Lesson Aggregator and Patch Aggregator independently synthesize repairs from the distilled lessons and candidate patches, respectively.
This allows the two aggregators to consider the agents' vulnerability diagnoses and concrete repair implementations separately.
Finally, the Aggregator reconciles their proposals, combining complementary changes and resolving conflicting assumptions to produce the final repair.

We evaluate \framework on 169 Python, JavaScript, and Go vulnerabilities from PatchEval~\cite{wei2025patcheval} and 35 C/C++ vulnerabilities from VulnLoc+~\cite{ye2026well}. 
On PatchEval, \framework achieves repair success rates of 39.6\% and 71.6\% with GPT-5.6-Luna and DeepSeek-V4.1-Flash, respectively, exceeding a single-agent baseline using the same model by 7.1 and 14.2 percentage points.
Our framework also outperforms EvoRepair, a state-of-the-art vulnerability repair approach, by 16.0 and 25.4 percentage points, respectively.
These results demonstrate that \framework's improvements are unlikely to be attributable to additional tools or instructions.
On VulnLoc+, \framework achieves  repair success rates of 51.4\% and 62.9\%, exceeding LoopRepair by 20.0 and 17.1 percentage points. 
Even when LoopRepair is given access to the ground-truth PoV, \framework outperforms it by 2.9 and 11.4 percentage points, respectively.

In short, this paper makes three contributions:
\begin{itemize}
  \item We formulate information-constrained vulnerability repair as a task of repairing the vulnerability using repository context without relying on vulnerability metadata or oracle feedback.
  \item We propose \framework, a new automated vulnerability repair approach that employs evidence-specialized exploration with hierarchical lesson–patch aggregation to guide repair generation.
  \item We evaluate \framework on two benchmarks comprising 204 vulnerabilities in total, showing that it substantially outperforms state-of-the-art vulnerability repair methods.
\end{itemize}

\ifdefined\mainmanuscript
\else
\bibliography{refs}
\end{document}
\fi

\ifdefined\mainmanuscript
\else
\documentclass[11pt]{article}
\usepackage[review]{acl}
\usepackage{times}
\usepackage{latexsym}
\usepackage[T1]{fontenc}
\usepackage[utf8]{inputenc}
\usepackage{microtype}
\usepackage{inconsolata}
\usepackage{graphicx}
\usepackage{xspace}
\usepackage{array}
\usepackage{multirow}
\usepackage{subcaption}
\usepackage{booktabs}
\usepackage{xcolor}
\graphicspath{{images/}{../images/}}
\newlength{\agentH}
\newlength{\agentWa}
\newlength{\agentWb}
\newlength{\agentWc}
\newcommand{\framework}{\textsc{ExploreRepair}\xspace}
\newcommand{\Description}[1]{}
\newcommand{\change}[1]{\textcolor{blue}{#1}}
\newcommand{\manh}[1]{}
\newcommand{\hj}[1]{}
\newcommand{\iv}[1]{}
\newcommand{\rw}[1]{}
\begin{document}
\fi

\begin{figure*}[!th]
  \centering
  \includegraphics[width=\textwidth]{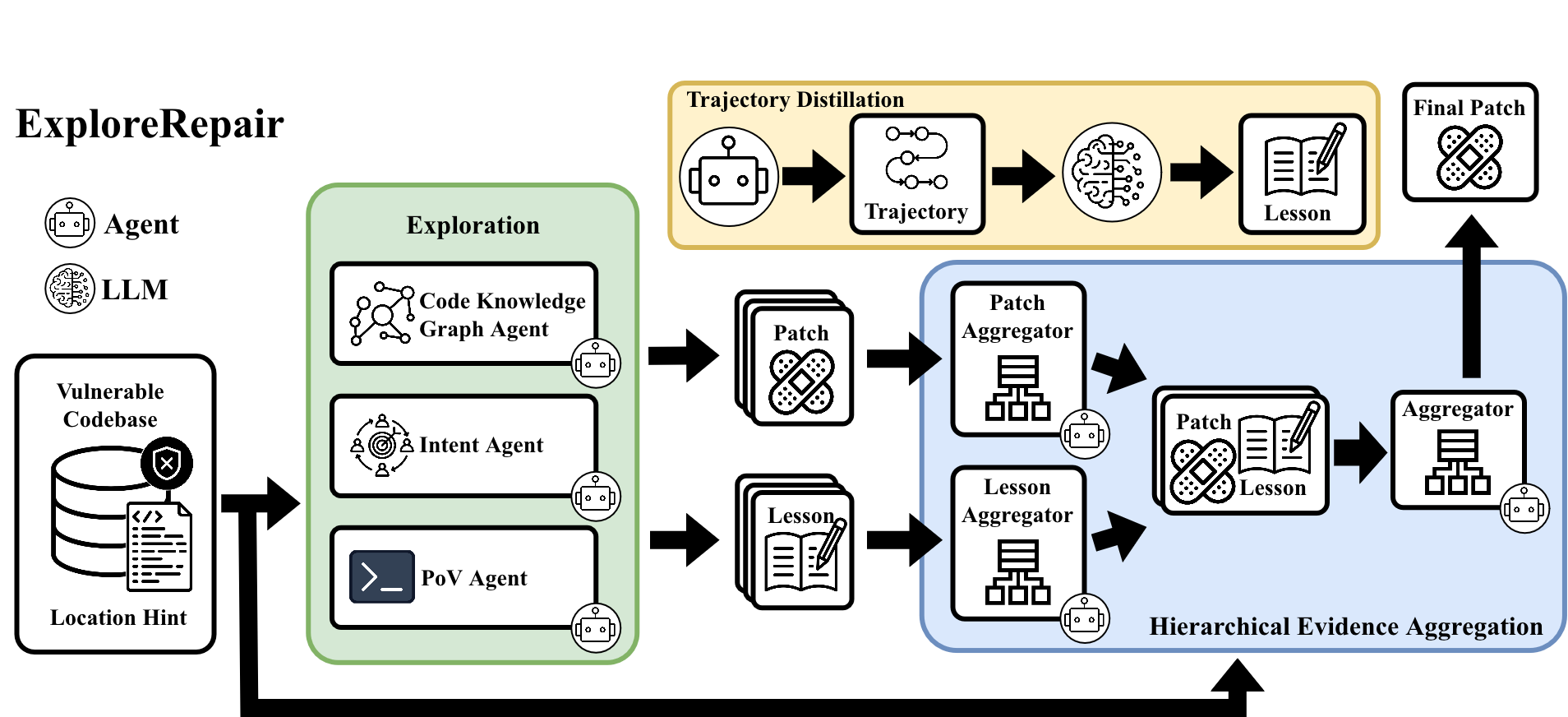}
  \caption{\framework has two phases. (1) In its exploration phase, three exploration agents independently produce trajectories and patches.
  (2) In its aggregation phase, the Lesson Aggregator and Patch Aggregator synthesize repairs from the agents' observations and repair implementations, respectively. The final Aggregator reconciles these two repairs.
  Trajectory distillation uses an LLM to summarize exploration trajectories and candidate patches into role-specific lessons, and aggregation trajectories into lessons for the final Aggregator.}
  \label{fig:overview}
\end{figure*}

\section{Related Work}
\label{sec:related}

\subsection{Automated Vulnerability Repair}

Recent program analysis-based approaches generate repairs from executions that trigger the vulnerability. 
They often require language-specific analyses or a demonstration of the security flaw~\cite{hong2020saver,gao2021beyond,zhang2022program,shariffdeen2025vulnerability}. 
For example, LoopRepair requires a PoV test and uses information from its execution to refine C/C++ patches~\cite{ye2026well}.

Training-based approaches learn from historical fixes using CWE knowledge, security patterns, or compiler feedback~\cite{zhou2024out,li2026vulkey,yang2026securepair,zhang2026synthfix}. 
Recent LLM-based repair systems use repository context~\cite{nong2025appatch,heo2026intentfix,rashid2026logiceval} to understand vulnerabilities. 
For example, IntentFix analyzes the codebase and reconstructs developer intent when synthesizing a repair.
EvoRepair~\cite{hu2026evorepair} retrieves experiences from earlier repairs, guided by ground-truth PoV tests.
ContraFix~\cite{liu2026contrafix} mutates the PoV to contrast runtime behavior.
KeaRepair~\cite{cao2026knowledge} combines verified program facts with knowledge from historical fixes.
AgenticRepair~\cite{fu2026agenticrepair} uses a multi-agent approach to examine different context sources.
Similar to \framework, these methods coordinate agents for codebase exploration; however, they do not prevent premature commitment to an incorrect repair.

\subsection{Multi-Agent Reasoning and Trajectory Distillation}

Multi-agent systems assign software tasks to specialized roles, as in MetaGPT and ChatDev~\cite{hong2024metagpt,qian2024chatdev,islam2024mapcoder}. Related systems coordinate agents for code generation, review, vulnerability detection, and general program repair~\cite{tang2024codeagent,luo2026unlocking,widyasari2026let,chen2025autoreview,lee2025unidebugger}.
While most organize agents around discussion of a shared task, 
\framework assigns each agent to explore a different dimension of the codebase independently.

While prior work used multi-agent approaches, such approaches do not automatically improve  diversity or correctness ~\cite{encheng2026single}. 
In particular, incorrect initial analysis can mislead other agents~\cite{oh2026belief,yoffe2025debunc,zhu2026demystifying}. To avoid this, \framework uses independent agents for exploration, with each agent specializing in a single evidence type.

An agent's trajectory is a sequence of thoughts, actions, and environment observations generated while solving a task~\cite{yao2023react}. Prior work summarizes information from trajectories into feedback, workflows, or reusable strategies~\cite{shinn2023reflexion,wang2025agent,yang-etal-2025-coarse}. ReasoningBank distills lessons from successful and failed trajectories for later retrieval~\cite{ouyang2026reasoningbank}.
Because each trajectory is lengthy and excessively large contexts can lead to confused reasoning~\cite{liu2024lost}, \framework distills each agent's exploration into lessons during aggregation, before synthesizing the repair.

\ifdefined\mainmanuscript
\else
\bibliography{refs}
\end{document}
\fi

\ifdefined\mainmanuscript
\else
\documentclass[11pt]{article}
\usepackage[review]{acl}
\usepackage{times}
\usepackage{latexsym}
\usepackage[T1]{fontenc}
\usepackage[utf8]{inputenc}
\usepackage{microtype}
\usepackage{inconsolata}
\usepackage{graphicx}
\usepackage{xspace}
\usepackage{array}
\usepackage{multirow}
\usepackage{subcaption}
\usepackage{booktabs}
\usepackage{xcolor}
\graphicspath{{images/}{../images/}}
\newlength{\agentH}
\newlength{\agentWa}
\newlength{\agentWb}
\newlength{\agentWc}
\newcommand{\framework}{\textsc{ExploreRepair}\xspace}
\newcommand{\Description}[1]{}
\newcommand{\manh}[1]{}
\newcommand{\hj}[1]{}
\newcommand{\iv}[1]{}
\newcommand{\rw}[1]{}
\begin{document}
\fi

\section{Problem Setting}
\label{sec:problem-statement}

Let $C$ denote the repository at its vulnerable revision and let $\ell$ denote a function-level localization hint containing one or more file and line ranges. 
Prior AVR methods may also receive additional vulnerability information, such as CWE labels or  developer-written PoV tests, as well as feedback from evaluating candidate patches against a benchmark oracle. 
We denote these artifacts by $F$.

In an \textbf{information-constrained vulnerability repair} setting, the repair agent $\mathcal{A}$ receives only $(C,\ell)$. It can inspect the repository's code and can execute its test suites, but has no access to $F$.
It produces a single patch $p$:
\[
p = \mathcal{A}(C,\ell).
\]
An ideal patch should eliminate the vulnerability, as determined by the benchmark's PoV test.
At the same time, it should preserve existing functionality, as assessed by the repository's unit tests~\cite{yu2026patch,rashid2026logiceval,sun2026swe}.


\ifdefined\mainmanuscript
\else
\bibliography{refs}
\end{document}
\fi

\ifdefined\mainmanuscript
\else
\documentclass[11pt]{article}
\usepackage[review]{acl}
\usepackage{times}
\usepackage{latexsym}
\usepackage[T1]{fontenc}
\usepackage[utf8]{inputenc}
\usepackage{microtype}
\usepackage{inconsolata}
\usepackage{graphicx}
\usepackage{xspace}
\usepackage{array}
\usepackage{multirow}
\usepackage{subcaption}
\usepackage{booktabs}
\usepackage{xcolor}
\graphicspath{{images/}{../images/}}
\newlength{\agentH}
\newlength{\agentWa}
\newlength{\agentWb}
\newlength{\agentWc}
\newcommand{\framework}{\textsc{ExploreRepair}\xspace}
\newcommand{\Description}[1]{}
\newcommand{\manh}[1]{}
\newcommand{\hj}[1]{}
\newcommand{\iv}[1]{}
\newcommand{\rw}[1]{}
\begin{document}
\fi

\section{\framework}
\label{sec:approach}

\begin{figure*}[!t]
  \centering
  \settowidth{\agentWa}{\includegraphics[height=100pt]{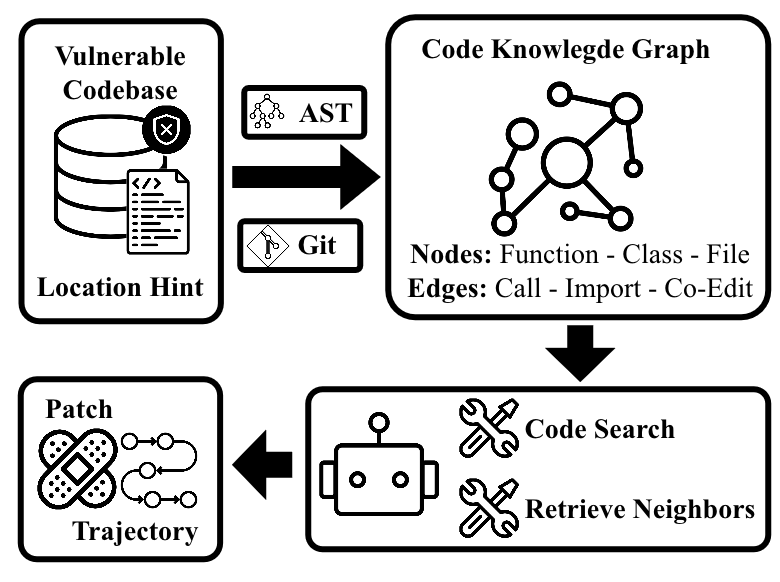}}%
  \settowidth{\agentWb}{\includegraphics[height=100pt]{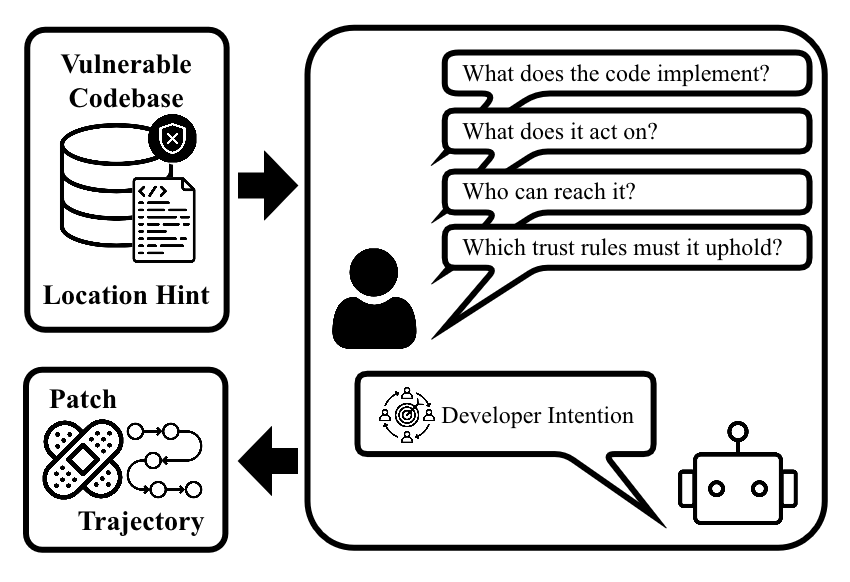}}%
  \settowidth{\agentWc}{\includegraphics[height=100pt]{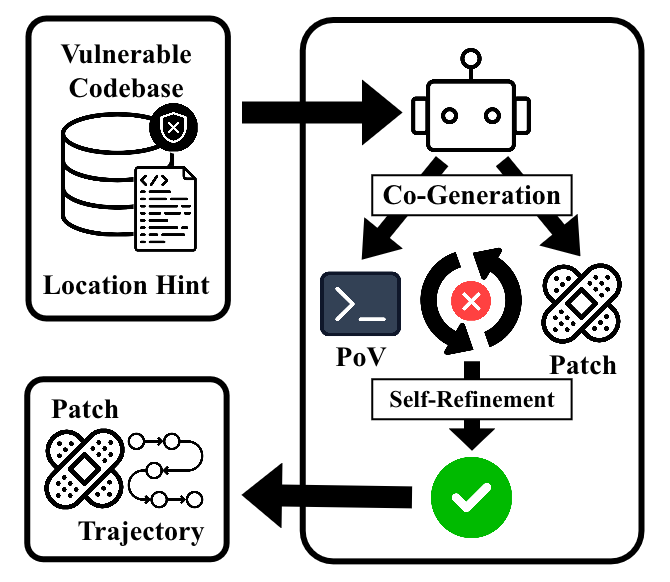}}%
  \setlength{\agentH}{\dimexpr 100pt * \numexpr\dimexpr\linewidth-2em\relax\relax / \numexpr\agentWa+\agentWb+\agentWc\relax\relax}%
  \settowidth{\agentWa}{\includegraphics[height=\agentH]{images/ckgraph.png}}%
  \settowidth{\agentWb}{\includegraphics[height=\agentH]{images/intent.png}}%
  \settowidth{\agentWc}{\includegraphics[height=\agentH]{images/poc.png}}%
  \begin{subfigure}[b]{\agentWa}
    \centering
    \includegraphics[height=\agentH]{images/ckgraph.png}
    \caption{Code Knowledge Graph}
    \label{fig:agent-ckg}
  \end{subfigure}\hfill
  \begin{subfigure}[b]{\agentWb}
    \centering
    \includegraphics[height=\agentH]{images/intent.png}
    \caption{Intent}
    \label{fig:agent-intent}
  \end{subfigure}\hfill
  \begin{subfigure}[b]{\agentWc}
    \centering
    \includegraphics[height=\agentH]{images/poc.png}
    \caption{PoV}
    \label{fig:agent-poc}
  \end{subfigure}
  \caption{The three exploration agents receive the same repository and localization hint. They investigate (a) repository structure, (b) intended behavior and trust constraints, and (c) failures that demonstrate the vulnerability.}
  \label{fig:context-agents}
\end{figure*}

\subsection{Overview}
\label{sec:overview}

Figure~\ref{fig:overview} outlines \framework, which is a two-phase approach.
First, in its exploration phase, \framework{} invokes three exploration agents that collect evidence along different dimensions of analysis.
Second, in the aggregation phase, the trajectories and patches produced by each exploration agent are distilled into lessons before they are aggregated into a single final patch, produced as the output of \framework.

In the exploration phase, each of the three exploration agents is individually tasked with producing a repair for the vulnerability.
To do so, each agent must collect and consider evidence along its assigned dimension of the codebase that may be relevant to the repair.
One agent collects structural evidence captured in a Code Knowledge Graph.
The graph represents files, functions, and classes and their call and import relationships, extracted from abstract syntax trees (ASTs). We also use Git history to link files changed in the same commit.
The Intent agent collects semantic evidence by focusing on reconstructing developer intent.
Finally, the PoV agent focuses on constructing executable evidence of the vulnerability. 
Each exploration agent produces a pair of a trajectory and a candidate patch. These outputs are used in the aggregation phase.

In the aggregation phase, to leverage the diverse information produced by the different evidence-specialized agents, the trajectories and patches are hierarchically aggregated.
First, we use an LLM to distill each agent's trajectory and candidate patch into a role-specific lesson covering code structure, developer intent, or observed execution, together with the proposed change and remaining uncertainties.
Next, in parallel, the Lesson Aggregator synthesizes a repair from the lessons, while the Patch Aggregator synthesizes one from the exploration agents' patches.
Each of these aggregators captures a complementary aspect of the repair. 
The former captures the agents' diagnoses and inferred security constraints, while the latter preserves concrete implementation details that may be lost during distillation.
The final Aggregator then reconciles the two proposals by examining their supporting evidence and implementation choices, combining compatible changes or resolving conflicting assumptions to produce the final repair.

\subsection{Evidence-Specialized Exploration}
\label{sec:exploration-agents}

\subsubsection{Structural Evidence}
\label{sec:graph-agent}

The Code Knowledge Graph agent investigates the code related to the localized function (Figure~\ref{fig:agent-ckg}). For each vulnerability, we construct a graph of functions, classes, and files from abstract syntax trees. Call edges connect callers and callees, import edges connect dependent files, and shared-commit edges connect files that developers changed together. We also embed the node text in a vector store for that vulnerability.

The agent uses two tools to query the graph. \texttt{search\_code} retrieves nodes related to a natural-language or code query, and \texttt{graph\_neighbors} returns calls and imports connected to a selected node. The agent traces how external inputs reach the localized function, examines helpers and validators, and compares related paths that perform similar operations. It uses these findings to diagnose the vulnerability and produce a patch, returning its trajectory alongside the patch.

\subsubsection{Semantic Evidence}
\label{sec:intent-agent}

The Intent agent reconstructs developer intent by identifying the expected behavior and trust assumptions before attempting a repair (Figure~\ref{fig:agent-intent}). It addresses four questions adapted from IntentFix~\cite{heo2026intentfix}: what specification or API contract governs the code, who can reach it and control its inputs, what resources it protects, and what constraints should hold for untrusted data. From these observations, the agent infers up to five security constraints that should be enforced.

The agent answers these questions in order and supports its claims with file and line references. It records uncertainty when the repository does not provide enough information. It then identifies where the implementation violates the inferred constraints and produces a patch, returning the trajectory that explains its diagnosis.

\subsubsection{Executable Evidence}
\label{sec:poc-agent}

The PoV agent constructs a test to demonstrate the vulnerability it has inferred (Figure~\ref{fig:agent-poc}). It first traces attacker-controlled input from a public API, command-line interface, or endpoint to the localized code. It then writes and executes a PoV test against the vulnerable repository, revising the test until the observed behavior demonstrates the suspected security violation. Finally, it repairs the root cause and reruns the same test to check the repair.

The agent derives this test from the repository and its own diagnosis. It cannot access the developer-written PoV used for benchmark evaluation. The prompt prohibits weakening the generated test or adding a repair specific to its input, and asks the agent to address the underlying flaw in production code. The agent returns its patch and the execution observations that support it.

\subsection{Hierarchical Evidence Aggregation}
\label{sec:aggregator}

The aggregation phase requires \framework to aggregate its findings while leveraging the complementary information from the trajectories and proposed patches.
Each agent provides a trajectory and a proposed patch.
Its trajectory provides evidence and justification for its diagnosis,
while its proposed patch provides implementation details, such as the code locations where new code should be added.

The Lesson Aggregator and Patch Aggregator first synthesize repairs separately, allowing each to examine the information in its inputs before a final Aggregator reconciles their proposals. 
This process has three steps.

\paragraph{1) Distilling the agents' observations.}
We use a separate LLM call to distill each agent's trajectory and candidate patch into a short lesson. The lesson records the suspected vulnerability, supporting code locations or execution observations, proposed change, and unresolved claims.
The LLM accounts for the agent's role: structural lessons focus on code paths and repair locations, semantic lessons focus on expected behavior and trust constraints, and executable lessons focus on the observed violation and the effect of the patch.

\paragraph{2) Aggregating observations and implementations separately.}
The Lesson Aggregator receives the three lessons and synthesizes a repair from the agents' observations. In parallel, the Patch Aggregator receives the three candidate diffs and synthesizes a repair from their concrete implementations. Both can inspect the original repository and are instructed to check the supplied attempts, since any of them may be incorrect. The lessons provide the reasoning and uncertainty behind a repair, while the patches retain implementation details that the lessons may omit.

\paragraph{3) Reconciling the repair proposals.}
After the Lesson Aggregator and Patch Aggregator propose their repairs, we use an LLM to distill each aggregator's trajectory into a lesson. The final Aggregator receives these two lessons and their corresponding patches, checks their claims against the repository, and produces the final patch.
It can adopt one repair, combine compatible changes, reject assumptions shared by both attempts, or construct a new patch. 
Its decision uses the supplied observations, implementation details, and repository code.

\ifdefined\mainmanuscript
\else
\bibliography{refs}
\end{document}
\fi

\ifdefined\mainmanuscript
\else
\documentclass[11pt]{article}
\usepackage[review]{acl}
\usepackage{times}
\usepackage{latexsym}
\usepackage[T1]{fontenc}
\usepackage[utf8]{inputenc}
\usepackage{microtype}
\usepackage{inconsolata}
\usepackage{graphicx}
\usepackage{xspace}
\usepackage{array}
\usepackage{multirow}
\usepackage{subcaption}
\usepackage{booktabs}
\usepackage{xcolor}
\graphicspath{{images/}{../images/}}
\newlength{\agentH}
\newlength{\agentWa}
\newlength{\agentWb}
\newlength{\agentWc}
\newcommand{\framework}{\textsc{ExploreRepair}\xspace}
\newcommand{\Description}[1]{}
\newcommand{\change}[1]{\textcolor{blue}{#1}}
\newcommand{\manh}[1]{}
\newcommand{\hj}[1]{}
\newcommand{\iv}[1]{}
\newcommand{\rw}[1]{}
\begin{document}
\fi

\section{Experimental Setup}
\label{sec:setup}

\subsection{Benchmark}
\label{sec:benchmark}

PatchEval provides runnable test artifacts and prebuilt Docker images for 230 CVEs across Python, JavaScript, and Go~\cite{wei2025patcheval}.
Of the 176 instances with both a PoV and a unit-test suite, we exclude seven whose benchmark test patches do not compile against the vulnerable repository.
We evaluate on the remaining 169 vulnerabilities: 64 in Python, 49 in JavaScript, and 56 in Go. Table~\ref{tab:patcheval-types} summarizes their vulnerability types.

To evaluate transfer to C/C++, we use VulnLoc+, the benchmark introduced with LoopRepair~\cite{ye2026well}.
We use 35 of the 40  vulnerabilities included in the benchmark, excluding vulnerabilities whose PoV tests already pass on the vulnerable version before a correct patch is applied.
VulnLoc+ provides PoVs but no unit tests.  Table~\ref{tab:vulnlocplus-types} summarizes the vulnerabilities in the benchmark.

\begin{table}[t]
  \caption{PatchEval CVEs by language and CWE. A CVE may have multiple CWEs; \emph{Other} groups the 27 types with fewer than 10 instances.}
  \label{tab:patcheval-types}
  \centering
  \footnotesize
  \setlength{\tabcolsep}{2pt}
  \begin{tabular*}{\linewidth}{@{\extracolsep{\fill}}l l rrr r@{}}
    \toprule
    \textbf{Vulnerability Type} & \textbf{CWE} & \textbf{Py} & \textbf{JS} & \textbf{Go} & \textbf{Total} \\
    \midrule
    Path traversal           & CWE-22  & 14 & 4 & 9 & 27 \\
    Code injection           & CWE-94  & 11 & 9 & 6 & 26 \\
    Ext.\ file path control  & CWE-73  & 14 & 4 & 7 & 25 \\
    Command injection        & CWE-77  & 11 & 7 & 6 & 24 \\
    OS command injection     & CWE-78  & 11 & 6 & 6 & 23 \\
    Open redirect            & CWE-601 &  6 & 6 & 4 & 16 \\
    Unnecessary privileges   & CWE-250 &  6 & 2 & 4 & 12 \\
    Improper privilege mgmt. & CWE-269 &  6 & 2 & 4 & 12 \\
    Improper authorization   & CWE-285 &  6 & 2 & 4 & 12 \\
    SQL injection            & CWE-89  &  4 & 3 & 4 & 11 \\
    Missing authorization    & CWE-862 &  - & 4 & 6 & 10 \\
    Other (27 CWEs)          &         & 24 & 25 & 29 & 78 \\
    \midrule
    \textbf{CVEs} & & \textbf{64} & \textbf{49} & \textbf{56} & \textbf{169} \\
    \bottomrule
  \end{tabular*}
\end{table}

\begin{table}[t]
  \caption{VulnLoc+ ~\cite{ye2026well} vulnerabilities by type. The benchmark spans 9 C projects.}
  \label{tab:vulnlocplus-types}
  \centering
  \footnotesize
  \setlength{\tabcolsep}{2pt}
  \begin{tabular*}{\linewidth}{@{\extracolsep{\fill}}l r r@{}}
    \toprule
    \textbf{Type} & \textbf{Projects} & \textbf{Total} \\
    \midrule
    Buffer overflow          & 9 & 23 \\
    Integer overflow         & 3 &  3 \\
    Divide by zero           & 2 &  3 \\
    Null-pointer dereference & 2 &  2 \\
    Data-type overflow       & 1 &  2 \\
    Use after free           & 1 &  2 \\
    \midrule
    \textbf{Vulnerabilities} & \textbf{9} & \textbf{35} \\
    \bottomrule
  \end{tabular*}
\end{table}

\subsection{Metrics}
\label{sec:metrics}

On PatchEval, a patch is considered \textbf{plausible} if it passes the benchmark PoV test and \textbf{successful} if it also passes the repository's unit tests. Let $N$ denote the number of evaluated CVEs, $S$ the number of successful repairs, and $P$ the number of plausible patches. We report \textbf{Repair Success Rate (RSR)} and \textbf{Overfitting Rate (OFR)} as percentages:
\[
\mathrm{RSR}=\frac{S}{N}\times100,\qquad
\mathrm{OFR}=\frac{P-S}{N}\times100.
\]
OFR measures the proportion of all evaluated CVEs whose patches pass the PoV but fail the repository's unit tests.
As VulnLoc+ does not provide unit tests,
we measure RSR using the benchmark PoV alone
and do not report OFR for this benchmark.

\subsection{Baselines}
\label{sec:baselines}

We compare \framework with IntentFix, which infers developer intent~\cite{heo2026intentfix}, and EvoRepair, which retrieves experiences from prior trajectories~\cite{hu2026evorepair}. We adapt EvoRepair to exclude ground-truth PoVs and benchmark evaluation feedback. On VulnLoc+, we evaluate LoopRepair, a C/C++ repair approach~\cite{ye2026well}, in its released PoV-assisted setting and a no-PoV variant. We also include a standalone LLM, Single Agent, and Multi-Agent. 
Baseline configurations are provided in Appendix~\ref{app:baseline-details}.

\subsection{Exploration and Aggregation Configurations}
\label{sec:aggregation-strategies}

We evaluate the 3-in-1 Agent to compare
independent exploration against combining all
three exploration mechanisms within a single agent.
We also compare four aggregation configurations
using the same exploration agents' outputs.

\paragraph{3-in-1 Agent.}
The \textit{3-in-1 Agent} extends the Single Agent baseline
with Code Knowledge Graph tools, Intent guidance,
and PoV-generation guidance. It uses all three
exploration mechanisms within a single trajectory
to investigate the vulnerability and produce a
repair. This allows us to compare joint exploration
within one agent against independent explorations
followed by hierarchical aggregation.

\paragraph{Patch and Lesson Aggregators.}
We also compare four aggregation configurations
using the same exploration agents' outputs.
The \textit{Patch Aggregator} receives only the exploration agents' candidate patches. 
The \textit{Lesson Aggregator} receives only the three lessons distilled from their observations.
\textit{Direct Lessons + Patches} gives one agent the three lessons and candidate patches together.
In the complete system, the Lesson Aggregator and Patch Aggregator independently produce repairs. The final Aggregator receives these repairs and their distilled lessons, then reconciles their proposals.

\subsection{General-Purpose Coding Agent}
\label{sec:codex-setup}

We also compare \framework with Codex, OpenAI's general-purpose coding agent~\cite{openai2026codex}, to assess whether specialized exploration and aggregation improve vulnerability repair beyond a general-purpose coding workflow.
Codex uses its native agent loop and receives the same repository and localization hint, with benchmark artifacts removed and external capabilities disabled.
We run Codex with GPT-5.6-Luna  and DeepSeek-V4.1-Flash.

\subsection{Implementation and Hardware}
\label{sec:implementation}

Our experiments use OpenAI GPT-5.6-Luna 
and DeepSeek-V4.1-Flash through OpenRouter.
We set a time limit for each phase to 1800\,s per vulnerability.
The Code Knowledge Graph agent uses \texttt{bge-large-en-v1.5} to embed code and retrieves $k=5$ locations.
We run the agents in isolated, network-disabled Docker containers with the repository at its vulnerable commit.
To prevent data leakage, we remove benchmark PoVs, reference patches, and grading scripts from the containers. 
Repository tests remain available.
Throughout the experiments, the agents do not have access to the benchmark evaluation code or results.
The experiments were run on a machine running Ubuntu 22.04.5, with an AMD EPYC 7763 processor and 251\,GB of RAM.

\ifdefined\mainmanuscript
\else
\bibliography{refs}
\end{document}
\fi

\ifdefined\mainmanuscript
\else
\documentclass[11pt]{article}
\usepackage[review]{acl}
\usepackage{times}
\usepackage{latexsym}
\usepackage[T1]{fontenc}
\usepackage[utf8]{inputenc}
\usepackage{microtype}
\usepackage{inconsolata}
\usepackage{graphicx}
\usepackage{xspace}
\usepackage{array}
\usepackage{multirow}
\usepackage{subcaption}
\usepackage{booktabs}
\usepackage{xcolor}
\graphicspath{{images/}{../images/}}
\newlength{\agentH}
\newlength{\agentWa}
\newlength{\agentWb}
\newlength{\agentWc}
\newcommand{\framework}{\textsc{ExploreRepair}\xspace}
\newcommand{\Description}[1]{}
\newcommand{\change}[1]{\textcolor{blue}{#1}}
\newcommand{\manh}[1]{}
\newcommand{\hj}[1]{}
\newcommand{\iv}[1]{}
\newcommand{\rw}[1]{}
\begin{document}
\fi

\section{Results}
\label{sec:results}

\subsection{Overall Repair Effectiveness}
\label{sec:overall-results}

We investigate the overall effectiveness of \framework. 
Table~\ref{tab:main-results} shows the experimental results of the baselines, individual exploration agents, and \framework on PatchEval. 

\begin{table}[t]
  \caption{Repair outcomes on 169 PatchEval vulnerabilities. Each entry reports Repair Success Rate (RSR) followed by Overfitting Rate (OFR) in parentheses.}
  \label{tab:main-results}
  \centering
  \small
  \setlength{\tabcolsep}{4pt}
  \begin{tabular*}{\columnwidth}{@{\extracolsep{\fill}}lrr@{}}
    \toprule
    & \multicolumn{2}{c}{\textbf{RSR~$\uparrow$ (OFR~$\downarrow$)}} \\
    \cmidrule(l){2-3}
    \textbf{Approach} & \textbf{GPT-5.6-Luna} & \textbf{DeepSeek-V4.1-Flash} \\
    \midrule
    IntentFix                & 4.7\% (1.2\%)   & 9.5\% (1.2\%) \\
    EvoRepair                & 23.7\% (3.0\%)  & 46.2\% (1.8\%) \\
    LLM                      & 10.7\% (1.2\%)  & 23.7\% (2.4\%) \\
    Multi-Agent     & 30.2\% (1.2\%)  & 55.6\% (1.2\%) \\
    Single Agent    & 32.5\% (0.6\%)  & 57.4\% (0.6\%) \\
    \midrule
    Codex           & 34.3\% (0.6\%)  & 43.8\% (0.6\%) \\
    \midrule
    Structural (Graph)               & 36.1\% (1.2\%)  & 55.6\% (0.0\%) \\
    Semantic (Intent)                & 29.6\% (0.0\%)  & 59.8\% (0.0\%) \\
    Executable (PoV)              & 33.7\% (0.0\%)  & 58.0\% (0.6\%) \\
    \midrule
    3-in-1 Agent    & 30.8\% (1.2\%)  & 53.3\% (0.6\%) \\
    \midrule
    \framework               & \textbf{39.6\% (0.0\%)} & \textbf{71.6\% (0.6\%)} \\
    \bottomrule
  \end{tabular*}
\end{table}

\paragraph{\framework outperforms EvoRepair under the same information constraints.}
EvoRepair retrieves experiences from prior trajectories to guide successive repair attempts~\cite{hu2026evorepair}. With the adaptation described in Section~\ref{sec:baselines}, it achieves an RSR of 23.7\% using GPT-5.6-Luna. \framework achieves 39.6\%, a gain of 16.0 percentage points (Table~\ref{tab:main-results}). EvoRepair's OFR is 3.0\%, compared with 0.0\% for \framework. With DeepSeek-V4.1-Flash, EvoRepair achieves an RSR of 46.2\%, compared with 71.6\% for \framework, a gain of 25.4 percentage points. Its OFR is 1.8\%, compared with 0.6\% for \framework.

\paragraph{\framework achieves the highest repair success rate with both models.}
With GPT-5.6-Luna, \framework achieves an RSR of 39.6\%, a gain of 7.1 percentage points compared with 32.5\% for Single Agent.
With DeepSeek-V4.1-Flash, its RSR is 71.6\%, compared with 57.4\% for Single Agent, a gain of 14.2 percentage points.
IntentFix achieves RSRs of 4.7\% and 9.5\%, yielding gains of 34.9 and 62.1 percentage points for \framework, respectively.
\framework's OFR is 0.0\% with GPT-5.6-Luna and 0.6\% with DeepSeek-V4.1-Flash, compared with 0.6\% for Single Agent with either model.

Multi-Agent achieves RSRs of 30.2\% with GPT-5.6-Luna and 55.6\% with DeepSeek-V4.1-Flash, below Single Agent and \framework with both models. 
The improvements of \framework over the multi-agent baseline suggest that generic subagent delegation does not achieve the benefits of evidence-specialized explorations and hierarchical aggregation.
These results show that the effectiveness of \framework goes beyond a single model. 

\paragraph{The strongest exploration agent differs between models.}
The Code Knowledge Graph agent
performs best with GPT-5.6-Luna, achieving an RSR
of 36.1\%, whereas the Intent agent performs best
with DeepSeek-V4.1-Flash, achieving 59.8\%.
Neither agent performs best with both models.
Full results are available in Appendix~\ref{app:aggregation-motivation}.

\paragraph{Aggregating independent explorations outperforms the 3-in-1 Agent.}
The 3-in-1 Agent, which has access to all three exploration mechanisms, achieves RSRs of 30.8\% with GPT-5.6-Luna and 53.3\% with DeepSeek-V4.1-Flash. \framework improves on these rates by 8.9 and 18.3 percentage points, respectively.
This suggests that agents struggle to correctly reason over the different evidence types. 
The results show the effectiveness of aggregating independent explorations of each dimension. 

\subsection{Aggregation Analysis}
\label{sec:aggregation-results}

To understand the contribution of the aggregation phase to the effectiveness of \framework, we conduct a detailed ablation analysis. 
In these experiments, we compare the Lesson Aggregator, Patch Aggregator, Direct Lessons + Patches, and 
the complete \framework using the same exploration agents' outputs.

\begin{table}[t]
  \caption{Effectiveness of different aggregation methods  on PatchEval. Each entry reports Repair Success Rate (RSR) followed by Overfitting Rate (OFR) in parentheses.}
  \label{tab:aggregation-configurations}
  \centering
  \small
  \setlength{\tabcolsep}{4pt}
  \begin{tabular*}{\columnwidth}{@{\extracolsep{\fill}}lrr@{}}
    \toprule
    & \multicolumn{2}{c}{\textbf{RSR~$\uparrow$ (OFR~$\downarrow$)}} \\
    \cmidrule(l){2-3}
    \textbf{Configuration} & \textbf{GPT-5.6-Luna} & \textbf{\shortstack{DeepSeek-V4.1-\\Flash}} \\
    \midrule
    Lesson Aggregator & 39.1\% (0.6\%) & 63.9\% (0.6\%) \\
    Patch Aggregator & 37.3\% (0.6\%) & 69.2\% (1.2\%) \\
    Direct Lessons + Patches & 37.3\% (0.6\%) & 67.5\% (0.6\%) \\
    \framework & \textbf{39.6\% (0.0\%)} & \textbf{71.6\% (0.6\%)} \\
    \bottomrule
  \end{tabular*}
\end{table}

\paragraph{Reconciling the two aggregators' repairs improves success rates.}
\framework achieves RSRs of 39.6\% and 71.6\%, exceeding both separate aggregators and Direct Lessons + Patches. 
Compared with the stronger
separate aggregator, the final Aggregator repairs
one additional vulnerability with GPT-5.6-Luna
and four with DeepSeek-V4.1-Flash, corresponding
to gains of 0.6 and 2.4 percentage points,
respectively.

\paragraph{Aggregation enables successful repairs even when all three exploration agents fail.}
On PatchEval, \framework successfully repairs 11 vulnerabilities with DeepSeek-V4.1-Flash and 3 vulnerabilities with GPT-5.6-Luna that none of the three exploration agents could repair individually.
These results demonstrate that aggregation can produce successful repairs by combining information from individually unsuccessful explorations.

\paragraph{The stronger aggregator differs between models.}
While \framework outperforms the baselines with both models, there are differences in the individual performance of each aggregator.
The Lesson Aggregator outperforms the Patch Aggregator with GPT-5.6-Luna, achieving an RSR of 39.1\% versus 37.3\%, while the Patch Aggregator performs better with DeepSeek-V4.1-Flash, achieving 69.2\% versus 63.9\%.
Direct Lessons + Patches achieves RSRs of 37.3\% and 67.5\%, respectively. 
With either model, giving the lessons and patches to one agent does not outperform the stronger of the two separate aggregators.

\subsection{C/C++ Programming Language Transfer}
\label{sec:transfer-results}

Next, we investigate whether \framework's effectiveness transfers to other programming languages.
Table~\ref{tab:vulnlocplus-results} compares \framework with LoopRepair on VulnLoc+. We run LoopRepair both with its required PoV and without it. 

\begin{table}[t]
  \caption{Repair Success Rate (RSR) on 35 VulnLoc+ C/C++ vulnerabilities. OFR cannot be measured because this benchmark has no unit tests. LoopRepair + PoV receives the ground-truth PoV while the other configurations do not.}
  \label{tab:vulnlocplus-results}
  \centering
  \footnotesize
  \setlength{\tabcolsep}{3pt}
  \begin{tabular*}{\columnwidth}{@{\extracolsep{\fill}}lrr@{}}
    \toprule
    \textbf{Approach} & \textbf{\shortstack{GPT-5.6-\\Luna}} & \textbf{\shortstack{DeepSeek-V4.1-\\Flash}} \\
    & \multicolumn{2}{c}{\textbf{RSR~$\uparrow$}} \\
    \midrule
    LoopRepair + PoV    & 48.6\% & 51.4\% \\
    LoopRepair (no PoV) & 31.4\% & 45.7\% \\
    Single Agent & 48.6\% & 57.1\% \\
    \framework          & \textbf{51.4\%} & \textbf{62.9\%} \\
    \bottomrule
  \end{tabular*}
\end{table}

\paragraph{The framework transfers to C/C++.}
Without access to a provided PoV, \framework achieves RSRs of 51.4\% with GPT-5.6-Luna and 62.9\% with DeepSeek-V4.1-Flash, compared with 31.4\% and 45.7\% for LoopRepair without a PoV, gains of 20.0 and 17.1 percentage points, respectively.
Even when LoopRepair receives the ground-truth PoV, \framework achieves higher RSRs, exceeding
it by 2.9 and 11.4 percentage points, respectively.

\subsection{Comparison with a General-Purpose Coding Agent}
\label{sec:codex-comparison}

Next, we compare \framework with Codex~\cite{openai2026codex} using GPT-5.6-Luna and DeepSeek-V4.1-Flash at medium reasoning effort. Both receive the same localization hint and repository, with external capabilities disabled. 
Codex uses its native agent loop, and \framework uses ReAct agents with the exploration and aggregation phases. As Table~\ref{tab:main-results} shows, \framework achieves an RSR of 39.6\%, compared with 34.3\% for Codex, a gain of 5.3 percentage points. Their OFRs are 0.0\% and 0.6\%, respectively. With DeepSeek-V4.1-Flash, \framework achieves an RSR of 71.6\%, compared with 43.8\% for Codex, a gain of 27.8 percentage points. Both have an OFR of 0.6\%.
Under the same model and vulnerability information, \framework's exploration and aggregation also outperform a general-purpose coding agent.

\ifdefined\mainmanuscript
\else
\bibliography{refs}
\end{document}
\fi

\ifdefined\mainmanuscript
\else
\documentclass[11pt]{article}
\usepackage[review]{acl}
\usepackage{times}
\usepackage{latexsym}
\usepackage[T1]{fontenc}
\usepackage[utf8]{inputenc}
\usepackage{microtype}
\usepackage{inconsolata}
\usepackage{graphicx}
\usepackage{xspace}
\usepackage{array}
\usepackage{multirow}
\usepackage{subcaption}
\usepackage{booktabs}
\usepackage{xcolor}
\graphicspath{{images/}{../images/}}
\newlength{\agentH}
\newlength{\agentWa}
\newlength{\agentWb}
\newlength{\agentWc}
\newcommand{\framework}{\textsc{ExploreRepair}\xspace}
\newcommand{\Description}[1]{}
\newcommand{\manh}[1]{}
\newcommand{\hj}[1]{}
\newcommand{\iv}[1]{}
\newcommand{\rw}[1]{}
\begin{document}
\fi

\section{Conclusion}
\label{sec:conclusion}

We propose \framework, which collects vulnerability observations and repair implementations from three exploration agents. 
A Lesson Aggregator and a Patch Aggregator independently synthesize repairs from these outputs, and a final Aggregator reconciles their proposals without benchmark feedback.
Experiments on PatchEval and VulnLoc+ show that \framework outperforms the evaluated baselines, including IntentFix, LoopRepair, other LLM-based approaches, and Codex.
We plan to expand the evaluation of \framework to a more diverse range of  programming languages in future work.

\ifdefined\mainmanuscript
\else
\bibliography{refs}
\end{document}
\fi

\ifdefined\mainmanuscript
\else
\documentclass[11pt]{article}
\usepackage[review]{acl}
\usepackage{times}
\usepackage{latexsym}
\usepackage[T1]{fontenc}
\usepackage[utf8]{inputenc}
\usepackage{microtype}
\usepackage{inconsolata}
\usepackage{graphicx}
\usepackage{xspace}
\usepackage{array}
\usepackage{multirow}
\usepackage{subcaption}
\usepackage{booktabs}
\usepackage{xcolor}
\graphicspath{{images/}{../images/}}
\newlength{\agentH}
\newlength{\agentWa}
\newlength{\agentWb}
\newlength{\agentWc}
\newcommand{\framework}{\textsc{ExploreRepair}\xspace}
\newcommand{\Description}[1]{}
\newcommand{\change}[1]{\textcolor{blue}{#1}}
\newcommand{\manh}[1]{}
\newcommand{\hj}[1]{}
\newcommand{\iv}[1]{}
\newcommand{\rw}[1]{}
\begin{document}
\fi

\section*{Limitations}
\label{sec:limitations}

First, \framework's results depend on the underlying LLM.
While \framework outperforms the baselines for the two models used in our experiments,  the strongest exploration agent differs between them.
As such, the experimental results are likely to vary with other models, prompts, or sampling seeds.
We ran the experiments three times at a sampling temperature of 1, with repair counts varying by at most one across runs.

Second, some baselines were designed to use vulnerability reports or PoVs. 
Removing these artifacts transforms the setting for which the baseline approaches were designed.
We preserved their original code where possible, and thoroughly assessed the implementation details.
During this process, we identified an evaluation flaw in IntentFix's replication package, which has been confirmed by its authors.

\section*{Ethical Considerations}

Automated repair can help maintainers respond to security flaws, but an incorrect patch may give developers unwarranted confidence in the repair.
Automated repair methods can also reveal exploitable behavior that can be used by malicious adversaries.
However, advances in offensive capabilities also increase the need to improve automated defensive tools.

Our experiments use vulnerabilities from public benchmarks  within isolated, network-disabled containers. 
Agents used in our experiments were prevented from communicating over the network and interacting with other deployed systems.
The replication package contains repair infrastructure and public benchmark outputs, but excludes credentials and previously undisclosed vulnerabilities.

\ifdefined\mainmanuscript
\else
\bibliography{refs}
\end{document}
\fi

\bibliography{refs}

\clearpage
\appendix
\ifdefined\mainmanuscript
\else
\documentclass[11pt]{article}
\usepackage[review]{acl}
\usepackage{times}
\usepackage{latexsym}
\usepackage[T1]{fontenc}
\usepackage[utf8]{inputenc}
\usepackage{microtype}
\usepackage{inconsolata}
\usepackage{graphicx}
\usepackage{xspace}
\usepackage{array}
\usepackage{multirow}
\usepackage{subcaption}
\usepackage{booktabs}
\usepackage{xcolor}
\graphicspath{{images/}{../images/}}
\newlength{\agentH}
\newlength{\agentWa}
\newlength{\agentWb}
\newlength{\agentWc}
\newcommand{\framework}{\textsc{ExploreRepair}\xspace}
\newcommand{\Description}[1]{}
\newcommand{\manh}[1]{}
\newcommand{\hj}[1]{}
\newcommand{\iv}[1]{}
\newcommand{\rw}[1]{}
\begin{document}
\fi

\section{Motivation for Hierarchical Evidence Aggregation}
\label{app:aggregation-motivation}

\begin{figure*}[t]
  \centering
  \includegraphics[width=\textwidth]{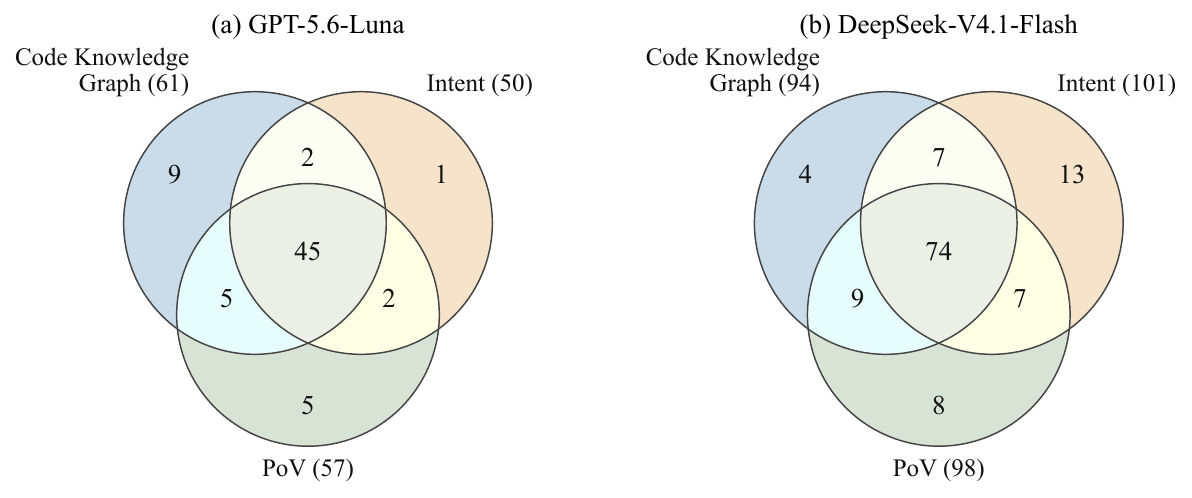}
  \caption{Overlap among successful repairs from the three evidence-specialized agents on 169 PatchEval vulnerabilities. Numbers indicate repairs achieved by exactly the agents whose circles intersect in that region; totals appear beside the agent names.}
  \label{fig:appendix-explorer-overlap}
\end{figure*}

We compare the vulnerabilities successfully repaired by each exploration agent to determine whether the agents repair different vulnerabilities.
The analysis uses the same 169 PatchEval vulnerabilities as Section~\ref{sec:overall-results}.
Each set contains the vulnerabilities whose patches pass both the benchmark PoV and unit tests. We count unsuccessful or missing attempts as failures.
Figure~\ref{fig:appendix-explorer-overlap} shows the overlap for both models.

\paragraph{Each agent repairs vulnerabilities that the others miss.}
With GPT-5.6-Luna, the Code Knowledge Graph agent achieves an RSR of 36.1\%, compared with 29.6\% for the Intent agent and 33.7\% for the PoV agent.
Their union covers 40.8\% of the evaluation set, including 4.7\% that the strongest agent misses.
Moreover, each role contributes successes that neither of the other two achieves: 5.3\% of the evaluation set for the Code Knowledge Graph agent, 0.6\% for the Intent agent, and 3.0\% for the PoV agent.
The same pattern holds with DeepSeek-V4.1-Flash: the union covers 72.2\% of the evaluation set, while the strongest agent, Intent, achieves an RSR of 59.8\%.
The Code Knowledge Graph, Intent, and PoV agents uniquely repair 2.4\%, 7.7\%, and 4.7\% of the evaluation set, respectively.
With both models, each agent successfully repairs some vulnerabilities that the other two do not. Using only the strongest agent would miss these repairs.

\paragraph{Aggregation must assess the attempts without benchmark feedback.}
These results motivate using all three agents' findings. However, selecting a successful candidate for each vulnerability would require knowing which patches pass the benchmark tests.
The agents do not receive those results, so aggregation must assess the observations behind each diagnosis and the implementation details in each patch.
\framework uses lessons to summarize the agents' vulnerability observations and patches to retain their concrete repair implementations. The Lesson Aggregator and Patch Aggregator independently synthesize repairs from these inputs, and a final Aggregator reconciles their proposals.
Section~\ref{sec:aggregation-results} evaluates this design. Appendix~\ref{app:aggregation-cases} examines cases in which aggregation produces a new fix or loses a successful one.

\ifdefined\mainmanuscript
\else
\bibliography{refs}
\end{document}
\fi

\ifdefined\mainmanuscript
\else
\documentclass[11pt]{article}
\usepackage[review]{acl}
\usepackage{times}
\usepackage{latexsym}
\usepackage[T1]{fontenc}
\usepackage[utf8]{inputenc}
\usepackage{microtype}
\usepackage{inconsolata}
\usepackage{graphicx}
\usepackage{xspace}
\usepackage{array}
\usepackage{multirow}
\usepackage{subcaption}
\usepackage{booktabs}
\usepackage{xcolor}
\graphicspath{{images/}{../images/}}
\newlength{\agentH}
\newlength{\agentWa}
\newlength{\agentWb}
\newlength{\agentWc}
\newcommand{\framework}{\textsc{ExploreRepair}\xspace}
\newcommand{\Description}[1]{}
\newcommand{\manh}[1]{}
\newcommand{\hj}[1]{}
\newcommand{\iv}[1]{}
\newcommand{\rw}[1]{}
\begin{document}
\fi

\section{Exploration Agents}
\label{app:explorer-details}

The exploration agents provide vulnerability observations and candidate patches for aggregation.
The Code Knowledge Graph, Intent, and PoV agents investigate repair scope, intended behavior, and observed failures, respectively.
This section gives further details of the agents described in Section~\ref{sec:exploration-agents} and how they gather evidence for their diagnosis.
All three use the same ReAct scaffold and independently investigate the same vulnerable revision. They receive the repository and localization hint under the setting defined in Section~\ref{sec:problem-statement}.
They can inspect files, execute shell commands and repository tests, and edit production code beyond the localized function.
Each agent runs at medium reasoning effort with a time limit of 1800\,s. It returns a production-code patch and a trajectory that records its observations.

\subsection{Code Knowledge Graph Agent}

The Code Knowledge Graph agent investigates where a repair should be applied and whether the same flaw appears along related paths.
Before editing, it traces callers toward entry points and external inputs, examines the helpers and validators used by the localized code, and compares sibling paths that perform similar operations.
This analysis can reveal that a guard belongs in a shared helper or that a local fix needs to cover another branch.

For each vulnerability, we construct a separate graph from the repository at its vulnerable revision, using Tree-sitter to parse source files into abstract syntax trees.
The graph contains file nodes and symbol nodes for functions, methods, classes, and other language-specific definitions. Symbol nodes retain qualified names and source locations so that the agent can inspect the corresponding code.
We resolve call sites and import statements to repository-local symbols and files to create call and import edges.
When commit history is available, we add edges between files changed in the same commit. We use only commits up to the vulnerable revision and weight each edge by the number of commits in which the two files changed together.

Two tools support this investigation.
\texttt{search\_code} retrieves locations from a per-instance vector index built with \texttt{bge-large-en-v1.5}, returning five results by default.
\texttt{graph\_neighbors} lists incoming and outgoing relations for a symbol or file, with a default limit of 20 neighbors per direction.
Both tools help locate relevant source code, while the prompt requires the agent to read that code through the shell before relying on it.
The agent records what the relationships imply about repair scope, then implements its patch.
We use an LLM to distill the agent's trajectory and proposed patch into a lesson that includes the repository relationships that led the agent to choose the repair location.

\subsection{Intent Agent}

The Intent agent establishes what the code is supposed to do before deciding how it is broken.
Its analysis follows the four intent questions described in Section~\ref{sec:intent-agent}, within a single repair trajectory.
It begins with the governing specification or API contract, looking for evidence in documentation, comments, tests, constants, and error messages.
It then traces the parties whose inputs reach the localized code and determines what the repository assumes about their trustworthiness.
Using those findings, it identifies the resources the code accesses and the checks already protecting them, including conventions followed by sibling paths.

The agent uses these findings to identify at most five constraints on untrusted input.
Each constraint must name the relevant values and operations, explain when it applies, and identify where it should be enforced.
The prompt requires file and line references and asks the agent to acknowledge claims that the repository does not establish.
Only after completing this analysis does it diagnose the implementation and produce a repair.
Its trajectory records the expected behavior and trust assumptions behind the patch. These observations help explain why the proposed checks are needed.

\subsection{PoV Agent}

The PoV agent tests whether its diagnosis corresponds to an observable security violation.
It first traces input from a public API, command-line interface, or endpoint to the localized code, then constructs a PoV through that entry point.
The test checks the program's behavior and must be executed against the vulnerable revision.
The agent revises it until the observed failure demonstrates the suspected violation.

Once the test provides that evidence, the agent repairs the production code and reruns the same test.
Its guidance asks it to address every site sharing the weakness and prohibits weakening the test or special-casing its input.
The agent saves generated tests in recognizable test paths. The harness uses these paths to exclude the tests from the production-code diff submitted for evaluation.
The trajectory records the input, the observed failure, and the response to the repair.
Passing the generated PoV supports the agent's diagnosis, but does not guarantee that its patch passes the hidden benchmark PoV.
During aggregation, the agents must therefore check these execution observations against the repository and the other agents' findings.

\ifdefined\mainmanuscript
\else
\bibliography{refs}
\end{document}
\fi

\ifdefined\mainmanuscript
\else
\documentclass[11pt]{article}
\usepackage[review]{acl}
\usepackage{times}
\usepackage{latexsym}
\usepackage[T1]{fontenc}
\usepackage[utf8]{inputenc}
\usepackage{microtype}
\usepackage{inconsolata}
\usepackage{graphicx}
\usepackage{xspace}
\usepackage{array}
\usepackage{multirow}
\usepackage{subcaption}
\usepackage{booktabs}
\usepackage{xcolor}
\usepackage{minted}
\graphicspath{{images/}{../images/}}
\newlength{\agentH}
\newlength{\agentWa}
\newlength{\agentWb}
\newlength{\agentWc}
\newcommand{\framework}{\textsc{ExploreRepair}\xspace}
\newcommand{\Description}[1]{}
\newcommand{\manh}[1]{}
\newcommand{\hj}[1]{}
\newcommand{\iv}[1]{}
\newcommand{\rw}[1]{}
\begin{document}
\fi

\section{Qualitative Aggregation Analysis}
\label{app:aggregation-cases}

We examine two cases in which aggregation produces a new fix or loses a correct one. Both cases use GPT-5.6-Luna.

\subsection{Producing a Fix from Unsuccessful Attempts}

CVE-2024-42005 affects how Django handles names supplied to a database query.
These names must be checked for dangerous characters; otherwise, an attacker can use a crafted name to change the meaning of the database command, causing SQL injection.
The affected function handles simple requests through a shortcut and more complex requests through a separate path (Listing~\ref{lst:django-vulnerable}).
The missing protection is the same in both: the names are used without the required safety check.

\begin{listing}[!t]
\begin{minted}[
frame=lines,
framesep=2mm,
fontsize=\scriptsize,
breaklines=true
]{python}
# fields contains the names requested by the user.
def set_values(self, fields):
    ...
    if fields:
        ...
        # Simple request: take the shortcut.
        if not self.extra and not self.annotations:
            # Copy the names without checking safety.
            field_names = list(fields)
        else:
            # More complex request: process each name.
            self.default_cols = False
            for f in fields:
                # f is one requested name.
                # No safety check here either.
                ...
    ...
\end{minted}
\caption{Original Django code: both paths lack the safety check. Dots stand for code omitted from the excerpt.}
\label{lst:django-vulnerable}
\end{listing}

Each exploration agent finds part of the problem, as shown in Listing~\ref{lst:django-specialized}.
The Code Knowledge Graph agent checks names on the complex path, but misses the shortcut.
The Intent agent adds a check in another function that creates query names.
The PoV agent adds \texttt{\#} to the list of forbidden characters, but does not make either path perform the check.
All three repairs fail the benchmark PoV test because the attack takes the unchecked shortcut.

\begin{listing}[!t]
\begin{minted}[
frame=lines,
framesep=2mm,
fontsize=\scriptsize,
breaklines=true,
highlightlines={7,14,23},
highlightcolor=blue!10
]{python}
# Code Knowledge Graph agent: attack test fails.
# This edit is on the complex path only.
else:
    self.default_cols = False
    for f in fields:
        # Reject a name containing forbidden characters.
        self.check_alias(f)
        # Remaining code omitted.

# Intent agent: attack test fails.
# Check a name created by a different function.
def add_filtered_relation(self,
                         filtered_relation, alias):
    self.check_alias(alias)
    filtered_relation.alias = alias
    ...

# PoV agent: attack test fails.
# This pattern lists forbidden characters.
# Add #, which can start an SQL comment.
# The pattern helps only where a check is called.
FORBIDDEN_ALIAS_PATTERN = _lazy_re_compile(
    r"['`\"\]\[;\s]|--|/\*|\*/|#"
)
\end{minted}
\caption{Specialized agents' Django repairs: each adds a useful protection, but none checks the shortcut used by the attack.}
\label{lst:django-specialized}
\end{listing}

The Lesson Aggregator uses these findings to inspect the function again and notices that the shortcut still has no check.
It keeps the earlier protections and adds a check for every name on the shortcut (Listing~\ref{lst:django-aggregated}).
The resulting repair passes both the benchmark PoV and unit tests.

\begin{listing}[!t]
\begin{minted}[
frame=lines,
framesep=2mm,
fontsize=\scriptsize,
breaklines=true,
highlightlines={16,17},
highlightcolor=blue!10
]{python}
# Keep the PoV agent's forbidden-character change.
FORBIDDEN_ALIAS_PATTERN = _lazy_re_compile(
    r"['`\"\]\[;\s]|--|/\*|\*/|#"
)

# Keep the Intent agent's check in this function.
def add_filtered_relation(self,
                         filtered_relation, alias):
    self.check_alias(alias)
    filtered_relation.alias = alias
    ...

# NEW: check names on the simple path too.
if not self.extra and not self.annotations:
    # Visit each requested name and check its safety.
    for f in fields:
        self.check_alias(f)
    # Use the names only after these checks.
    field_names = list(fields)
else:
    self.default_cols = False
    for f in fields:
        # Keep the Code Knowledge Graph agent's check.
        self.check_alias(f)
        # Remaining code omitted.
\end{minted}
\caption{Successful aggregated Django repair. The highlighted lines add the missing shortcut check, which no specialized agent produced.}
\label{lst:django-aggregated}
\end{listing}

Combining the exploration agents' patches leaves the shortcut unprotected, as does the Patch Aggregator's repair.
The Lesson Aggregator supplies the missing check, which the final Aggregator retains.
The final repair passes both the benchmark PoV and unit tests, even though all three exploration agents' repairs fail.

\subsection{Losing a Correct Fix During Aggregation}

CVE-2021-39163 affects how Synapse lists rooms associated with a group.
Group membership does not imply membership in each room, so a group member must not receive private-room metadata solely because the room is listed in the group.
The original code passes \texttt{allow\_private=True} when generating room entries, bypassing the private-room visibility check (Listing~\ref{lst:synapse-vulnerable}).

\begin{listing}[!t]
\begin{minted}[
frame=lines,
framesep=2mm,
fontsize=\scriptsize,
breaklines=true
]{python}
# Group visibility and room visibility are distinct.
room_results = await self.store.get_rooms_in_group(
    group_id, include_private=is_user_in_group
)
chunk = []
for room_result in room_results:
    room_id = room_result["room_id"]
    joined_users = await self.store.get_users_in_room(room_id)
    entry = await self.room_list_handler.generate_room_entry(
        room_id, len(joined_users),
        with_alias=False, allow_private=True
    )
    if not entry:
        continue
    # Remaining entry fields omitted.
    chunk.append(entry)
return {"chunk": chunk,
        "total_room_count_estimate": len(room_results)}
\end{minted}
\caption{Original Synapse listing: private-room metadata is allowed regardless of room membership.}
\label{lst:synapse-vulnerable}
\end{listing}

The Code Knowledge Graph agent checks whether the requester belongs to the room and counts only the returned entries (Listing~\ref{lst:synapse-specialized}).
Its repair passes both the benchmark PoV and unit tests.
The Intent and PoV agents instead use group membership to control private-room visibility; both repairs fail the benchmark PoV.

\begin{listing}[!t]
\begin{minted}[
frame=lines,
framesep=2mm,
fontsize=\scriptsize,
breaklines=true,
highlightlines={2,5,8,15},
highlightcolor=blue!10
]{python}
# Code Knowledge Graph agent: PoV and unit tests pass.
allow_private = requester_user_id in joined_users
entry = await self.room_list_handler.generate_room_entry(
    room_id, len(joined_users), with_alias=False,
    allow_private=allow_private,
)
# Remaining loop omitted; count only returned rooms.
return {"chunk": chunk, "total_room_count_estimate": len(chunk)}

# Intent and PoV agents: benchmark PoV fails.
# Both use group membership in the listing and summary.
entry = await self.room_list_handler.generate_room_entry(
    room_id, len(joined_users),
    with_alias=False,
    allow_private=is_user_in_group,
)
\end{minted}
\caption{Exploration agents' Synapse repairs: only the Code Knowledge Graph agent uses room membership as the authorization boundary.}
\label{lst:synapse-specialized}
\end{listing}

The Lesson Aggregator adopts the group-membership interpretation.
The Patch Aggregator also selects it, explicitly rejecting the room-membership check as inconsistent with the group API's visibility semantics and omitting the count correction.
The final Aggregator accepts this shared interpretation and produces the same unsuccessful changes (Listing~\ref{lst:synapse-aggregated}).

\begin{listing}[!t]
\begin{minted}[
frame=lines,
framesep=2mm,
fontsize=\scriptsize,
breaklines=true,
highlightlines={5,9},
highlightcolor=blue!10
]{python}
# Both intermediate aggregators and the final Aggregator
# use this check in the listing and summary paths.
entry = await self.room_list_handler.generate_room_entry(
    room_id, len(joined_users), with_alias=False,
    allow_private=is_user_in_group,
)
# The listing still counts rooms before filtering.
return {"chunk": chunk,
        "total_room_count_estimate": len(room_results)}
\end{minted}
\caption{Unsuccessful final Synapse repair: group membership still permits disclosure of private-room metadata.}
\label{lst:synapse-aggregated}
\end{listing}

The final repair fails the benchmark PoV: a group member who has not joined a private room still receives its name and metadata.
This case shows that the complete aggregation phase can discard a successful exploration repair when both intermediate aggregators adopt the same incorrect authorization assumption.

\ifdefined\mainmanuscript
\else
\bibliography{refs}
\end{document}
\fi

\ifdefined\mainmanuscript
\else
\documentclass[11pt]{article}
\usepackage[review]{acl}
\usepackage{times}
\usepackage{latexsym}
\usepackage[T1]{fontenc}
\usepackage[utf8]{inputenc}
\usepackage{microtype}
\usepackage{inconsolata}
\usepackage{xspace}
\usepackage{xcolor}
\newcommand{\framework}{\textsc{ExploreRepair}\xspace}
\newcommand{\change}[1]{\textcolor{blue}{#1}}
\begin{document}
\fi

\section{Baseline Configurations and Exclusions}
\label{app:baseline-details}

\paragraph{IntentFix.}
IntentFix infers developer intent from the vulnerable file to generate and refine a patch~\cite{heo2026intentfix}.

\paragraph{EvoRepair.}
EvoRepair retrieves experiences from prior trajectories to guide successive repair attempts~\cite{hu2026evorepair}. Its original implementation accesses the ground-truth PoV and benchmark evaluation feedback; we hide both in our experiments. The original study allows up to 15 repair turns, whereas our adaptation allows up to three refinement turns under the time limit described in Section~\ref{sec:implementation}.

\paragraph{LoopRepair.}
LoopRepair is designed for C/C++ and uses PoVs and taint traces to rank and refine LLM patches~\cite{ye2026well}. We evaluate both its released PoV-assisted setting and a no-PoV variant on VulnLoc+. In the no-PoV variant, we replace CrashRepair's PoV-based localization with the same reported crash location provided to \framework and select the first generated patch that compiles, without PoV validation or iterative refinement.

\paragraph{LLM-based baselines.}
The standalone large language model (LLM) receives only the code at the vulnerable location and returns replacement functions in one response. It cannot inspect other files, use tools, or receive feedback. Single Agent uses a ReAct scaffold with shell tools and medium reasoning effort. Multi-Agent extends this scaffold with a \texttt{spawn\_subagent} tool, allowing a lead agent to create subagents of itself and assign subtasks to them. The lead agent uses their reports and code changes to produce the final repair.

\paragraph{Excluded approaches.}
We select baselines that support the target languages and can run with the vulnerability information available in our setting. We exclude training-based approaches, such as SeCuRepair~\cite{yang2026securepair} and VulMaster~\cite{zhou2024out}, which learn from reference fixes, because \framework is an inference-time approach. We attempted to evaluate APPATCH using its
released implementation, but could not run
it successfully in our environment. 
ContraFix requires a PoV test and therefore cannot be used in our setting~\cite{liu2026contrafix}. AgenticRepair and KeaRepair had no publicly available replication packages at the time of evaluation~\cite{fu2026agenticrepair,cao2026knowledge}.

\ifdefined\mainmanuscript
\else
\bibliography{refs}
\end{document}
\fi

\end{document}